\documentclass[a4paper,11pt]{article}
\pdfoutput=1

\usepackage{array}
\usepackage{amsfonts}
\usepackage{amsmath}
\usepackage{amssymb}
\usepackage[small]{caption}
\usepackage{cite}
\usepackage[top=1.2in, bottom=1.2in, left=1.1in, right=1.1in]{geometry}
\usepackage{graphicx,subfigure}
\graphicspath{{./}{./figures/}}
\usepackage{pifont}
\usepackage{slashed}
\usepackage{xcolor}
\usepackage{ulem}

\definecolor{mediumblue}{rgb}{0,0,0.8}
\usepackage{hyperref}
\hypersetup{
  linktocpage=true,
  colorlinks=true,
  citecolor=mediumblue,
  filecolor=mediumblue,
  linkcolor=mediumblue,
  urlcolor=mediumblue,
}
\newcommand{\mailref}[1]{\href{mailto:#1}{#1}}

\allowdisplaybreaks
\def\l{\ensuremath{\left}}
\def\r{\ensuremath{\right}}

\newcommand{\I}{\ensuremath{i\mkern1mu}}

\newcommand{\mc}{\mathcal}
\newcommand{\nno}{\nonumber}
\newcommand{\fr}{\frac}

\begin{document}

\begin{titlepage}
\def\thefootnote{\fnsymbol{footnote}}

\begin{flushright}
  \texttt{}\\
  \texttt{}
\end{flushright}

\begin{centering}
\vspace{0.5cm}
{\Large \bf \boldmath
  Phenomenology of a two-component dark matter model}

\bigskip

\begin{center}
  {\normalsize % \sffamily \bfseries
    Yeong~Gyun~Kim$^{a,}$\footnote{\mailref{ygkim@gnue.ac.kr}},
    Kang~Young~Lee$^{b,}$\footnote{\mailref{kylee.phys@gnu.ac.kr}}, and
    Soo-hyeon~Nam$^{c,}$\footnote{\mailref{glvnsh@gmail.com}}
    }\\[0.5cm]
  \small
  $^a${\em Department of Science Education, Gwangju National
    University of Education,\\
    Gwangju 61204, Republic of Korea}\\[0.1cm]
  $^b${\em Department of Physics Education \& Research Institute of Natural Science,\\
  	Gyeongsang National University, Jinju 52828, Republic of Korea}\\[0.1cm]
  $^c${\em Department of Physics, Korea University, Seoul 02841, Republic of Korea}

\end{center}
\end{centering}

\medskip

\begin{abstract}
  \noindent
  We study a two-component dark matter model 
  consisting of a Dirac fermion and a complex scalar 
  charged under new U(1) gauge group in the hidden sector. 
  The dark fermion plays the dominant component of dark matter 
  which explains the measured DM relic density of the Universe. It has no direct coupling to ordinary
  standard model particles, thus evading strong constraints from the direct DM detection experiments.
  The dark fermion is self-interacting through the light dark gauge boson and 
  it would be possible to address that this model can be a resolution to
 	the small scale structure problem of the Universe. 
  The light dark gauge boson, which interacts with the standard model sector, 
  is also stable and composes the subdominant DM component.
  We investigate the model parameter space allowed by current experimental constraints 
  and phenomenological bounds. We also discuss the sensitivity of future experiments 
  such as SHiP, DUNE and ILC, for the obtained allowed parameter space.
  
\end{abstract}

\vspace{0.5cm}

\end{titlepage}

% -------------------------------------------------------------------------
\renewcommand{\thefootnote}{\arabic{footnote}}
\setcounter{footnote}{0}

\setcounter{tocdepth}{2}
\noindent \rule{\textwidth}{0.3pt}\vspace{-0.4cm}\tableofcontents
\noindent \rule{\textwidth}{0.3pt}

\section{Introduction\label{sec:intro}}

We have compelling evidence for the existence of non-baryonic dark matter (DM) in the Universe. 
But, its physical properties remain one of the main mysteries in particle physics today.
Weakly Interacting Massive Particles (WIMPs) have been leading DM candidates, since they
can provide a right amount of DM relic density with the weak scale masses and couplings in a natural way
\cite{Bertone, Bauer, Profumo}.
However, several experimental attempts have failed to find clear evidence for WIMPs so far. 
The sensitivity of the DM direct detection experiments\cite{LUX, PandaX, XENON1T} is now approaching, 
without positive DM signals, $neutrino\,\, floor$ level which represents an almost irreducible background 
from coherent neutrino-nucleus scattering\cite{Monroe, Vergados, Gutlein, Billard, OHare}.

On the other hand, the collisionless cold DM (CDM) paradigm describes 
the large-scale structure of the Universe remarkably well. 
However, it suffers from the long-standing small-scale structure problems\cite{Tulin}.
For instance, while the mass density profile for CDM halos in simulations 
turns out to steeply increase toward galactic center at small radii\cite{Dubinski, NFW1, NFW2}, 
the rotation curves of many observed galaxies show flat central DM density profiles\cite{Flores, Moore, Burkert, McGaugh}.
In order to solve the problems such as the core-cusp problem, 
DM self-scattering cross section\footnote{We'd like to mention that the 'problem' might be solved by improved modeling of baryonic feedback 
on the clustering process without invoking large DM self-scattering cross section.} typically has to be of order 
$\sigma / m_{DM} \sim 1\, {\rm cm^2 \,g^{-1}}$ \cite{Spergel}. 
One way to have such a large cross section is to introduce a light mediator in a weakly coupled theory. 
In this case, the self-scattering cross section becomes large enough at small DM velocities 
through Sommerfeld enhancement, so that the small scale problems can be solved.
However, if the light mediator is unstable and decays to the standard model (SM) particles, 
a very strong constraint comes from the measurements of the Cosmic Microwave Background (CMB) radiation\cite{Galli} and
it basically rules out\cite{Bringmann} such models for solving the small scale problems.  
Therefore, the strong constraints from the CMB data require that the light mediator should be stable.
 
In this paper we study a two-component DM model\cite{Ma}, where two species of DM are introduced. 
One species (a dark Dirac fermion $\psi$) constitutes a dominant DM component. 
It has no direct coupling to the SM particles so that the direct DM searches are insensitive to the species,
thus avoiding the strong constraints from the direct DM search experiments.
The self-interaction of the dark fermion DM is mediated by a dark gauge boson $A_X$, which is so light
that the DM self-scattering cross section at present is large enough for addressing the small scale problems. 
The dark gauge boson $A_X$ itself is stable and constitutes a subdominant DM in the Universe.
Therefore, the annihilation process $\psi\bar\psi \rightarrow A_X A_X$ 
at the recombination epoch does not affect the CMB data.
We will investigate the model parameter space, 
allowed by various experimental and phenomenological constraints, 
and discuss the prospects for finding new physics signals from future experiments.

This paper is organized as follows. In section 2, we introduce the two component DM model.
In section 3, we consider various experimental and phenomenological constraints on the parameter space 
of the DM model. We also discuss the sensitivity of the ILC, SHiP and DUNE experiments for the allowed parameter space.
We summarize our conclusions in section 4.

%%%%%
\section{A two-component DM model}
\label{sec:model}

In this section, we introduce a two component dark matter model\cite{Ma}. 
The dark sector is composed of a complex scalar field $S$ and a Dirac
fermion field $\psi$, both of which are charged under a new dark $U(1)_X$ gauge group
and singlet under the SM gauge group.
The Lagrangian for the dark sector is given by the following
renormalizable interactions,
%--
\begin{align}
  \mc{L}^{\rm dark} = \bar\psi (\I \slashed{D}
   - m_{\psi}) \psi + (D_\mu S)^* D^\mu S  - m_s^2 S^* S - 
   \lambda_s (S^* S)^2 -\fr{1}{4}F_{\mu\nu}F^{\mu\nu}.
  \label{eq:lagrangian}
\end{align}
The covariant derivative is
\begin{align}
 D^\mu = \partial^\mu + i g_X A_{X}^\mu X,
\end{align}
where $X$ is the hidden $U(1)_X$ charge operator, and 
$A_X^\mu$ the corresponding gauge field.
Due to the imposition of dark charge conjugation symmetry,
there is no kinetic mixing term between $A_{X}$ and the $U(1)_Y$ gauge boson of the SM.

The Higgs portal interaction connects the dark sector and the SM sector, 
and the full scalar potential of the model is given as
\begin{align}
   V = m_h^2 H^\dagger H + \lambda_h (H^\dagger H)^2
     +\lambda_{hs} (H^\dagger H)(S^* S)
     +m_s^2 S^* S + \lambda_s (S^* S)^2.
\end{align}
%
%The SM Higgs potential is given as
%
%\begin{align}
%  V_\mathrm{SM} = -\mu_h^2 H^\dagger H + \lambda_h (H^\dagger H)^2 .
%\end{align}
%
The Higgs doublet $H$ is written in the unitary gauge after the
electroweak symmetry breaking as follows:
\begin{equation}
  H = \fr{1}{\sqrt{2}}
  \begin{pmatrix}
    0 \\ v_h + h
  \end{pmatrix}
\end{equation}
with $v_h \simeq 246$~GeV.
The singlet scalar field also develops a nonzero vacuum expectation
value, $v_s$, and the singlet scalar field is written as 
\begin{equation}
	S = \fr{1}{\sqrt{2}} (v_s + s).
\end{equation}
The mass parameters $m_h^2$ and $m_s^2$ can be expressed in terms of
other parameters by using the minimization condition of the full
scalar potential $V$, \textit{i.e.},
\begin{align}
  -m_h^2 &= \lambda_h v_h^2 + \frac{1}{2} \lambda_{hs} v_s^2 ,\nno\\
  -m_s^2 &= \lambda_s v_s^2 + \frac{1}{2} \lambda_{hs} v_h^2 .
\end{align}
The mass terms of the scalar fields are
\begin{equation}
  -\mc{L}_\mathrm{mass} =
  \fr{1}{2} \mu_h^2 h^2 + \fr{1}{2} \mu_s^2 s^2 + \mu_{hs}^2 h s
  \label{eq:higgs_mass_matrix}
\end{equation}
where
\begin{align}
  \mu_{h}^2 &= 2\lambda_h v_h^2 ,\nno\\
  \mu_{s}^2 &= 2\lambda_s v_s^2 ,\nno\\
  \mu_{hs}^2 &= \lambda_{hs} v_h v_s .
  \label{eq:baremass}
\end{align}
A non-vanishing value of $\mu_{hs}^2$ induces mixing between the SM Higgs field configuration $h$ 
and the singlet scalar field $s$ as
\begin{equation}
  \begin{pmatrix}
    h_1 \\ h_2
  \end{pmatrix} = \begin{pmatrix}
    \cos\theta_s & \sin\theta_s\\
    -\sin\theta_s & \cos\theta_s
  \end{pmatrix} \begin{pmatrix}
    h \\ s
  \end{pmatrix},
\end{equation}
where the mixing angle $\theta_s$ is given by
\begin{equation}
  \tan\theta_s = \fr{y}{1 + \sqrt{1 + y^2}},
\end{equation}
with $y \equiv 2\mu_{hs}^2 / (\mu_h^2 - \mu_s^2)$.
Then, the physical Higgs boson masses are
\begin{equation}
  m_{h_1,\,h_2}^2 = \fr{1}{2} \l[ (\mu_h^2 + \mu_s^2) \pm (\mu_h^2 -
  \mu_s^2) \sqrt{1 + y^2} \r],
\end{equation}
where $h_1$ is the SM-like Higgs boson with $m_{h_1} =
125$~GeV and $h_2$ is the singlet-like scalar boson.
Scalar cubic interaction terms relevant for our study are given by
\begin{eqnarray}
	V \supset c_{122}\, h_1 h_2^2 + c_{222}\, h_2^3 + ... ,
\end{eqnarray}
with
\begin{eqnarray}
	c_{122} &=& \lambda_h v_h (3 c_\theta s_\theta^2) + \lambda_s v_s (3 c_\theta^2 s_\theta) + 
	\frac{1}{2} \lambda_{hs} v_h (c_\theta^3- 2 c_\theta s_\theta^2) +
	\frac{1}{2} \lambda_{hs} v_s (-2 c_\theta^2 s_\theta + s_\theta^3), \\
	c_{222} &=& \lambda_h v_h (-s_\theta^3)  + \lambda_s v_s (c_\theta^3) +
	\frac{1}{2} \lambda_{hs} v_h (-c_\theta^2 s_\theta) +
	\frac{1}{2} \lambda_{hs} v_s (c_\theta s_\theta^2).
\end{eqnarray}

The dark gauge boson mass and interactions with the scalar particles, 
assuming $U(1)_X$ charge $X=1$ for the dark scalar, are obtained from
\begin{equation}
	(D_\mu S)^* D^\mu S =
	\fr{1}{2}(\partial_\mu s) (\partial^\mu s) +  
	\fr{1}{2} g_X^2 v_s^2 (A_X^\mu)^2 + g_X^2 v_s (s A_{X\mu} A_X^\mu) +
	\fr{1}{2} g_X^2 (s^2 A_{X\mu} A_X^\mu),
\end{equation}
which gives the dark gauge boson mass $m_{A_X} = g_X v_s$.

%In this model, we have two DM candidates. 
%The dark fermion $\psi$ is stable due to dark fermion number conservation. Therefore, it is a DM candidate.
%If $m_{A_X} < 2 m_\psi$, the dark gauge boson $A_X$ cannot decay to $\bar\psi\psi$ and is also stable, providing 
%another DM candidate.
%Here, we assume $A_X$ is also a DM candidate and adopt free model parameters as follows: the singlet-like scalar mass $m_{h_2}$, 
%the scalar mixing angle ${\rm sin}\theta_s$, the dark gauge boson mass $m_{A_X}$, 
%the dark gauge coupling $g_X$, in addition to the dark fermion mass $m_\psi$ and 
%the dark U(1) charge $X_\psi$ of the dark fermion.

%%%%%
\section{Phenomenology}
\label{sec:phenomenology}

\subsection{Dark matter phenomenology}

In this model, we have two DM candidates and one new scalar particle $h_2$ in addition to the SM particles.
The dark fermion $\psi$ is stable due to dark fermion number conservation. Therefore, it is a DM candidate.
If $m_{A_X} < 2\, m_\psi$, the dark gauge boson $A_X$ cannot decay to the dark fermion pair $\psi\bar\psi$ and is also stable 
thanks to dark charge conjugation symmetry, thus providing another DM candidate.
In this work, we assume $A_X$ is much lighter than the dark fermion, keeping in mind the small-scale problems. 
Accordingly, the dark gauge boson also becomes a DM candidate. 

The dark sector communicates with the SM sector through the Higgs portal interaction.
The new scalar $h_2$ is in thermal equilibrium with the SM particles if the interaction rate is larger than the expansion rate of the universe. 
The annihilation amplitute of $h_2 h_2 \rightarrow b\bar{b}$ is proportional to 
$\lambda_{hs} m_b / m_{h_1}^2$ so that the thermal equilibrium condition reads as \cite{Ma}
\begin{eqnarray}
	\frac{\lambda_{hs}^2 m_b^2}{m_{h_1}^4} T^3 > \frac{T^2}{m_{\textrm{Planck}}}.
\end{eqnarray}
At $T \sim m_\psi \sim 100$ GeV, it requires that $\lambda_{hs} > 10^{-7}$ for
the thermal equilibrium. 
In turn, it implies that $\rm sin\theta_s > 3.2 \times 10^{-9}$ for $m_{h_2} = 10$ MeV 
and $v_s = 2$ GeV. As we will see in the next section, the constraint that $h_2$ lifetime
should be smaller than 1 second requires larger $\sin\theta_s$ value 
(therefore, larger $\lambda_{hs}$ value) so that the thermal equilibrium condition is fulfilled.
In this work, we require it is the case. Then the dark sector would be in thermal equilibrium with the SM sector at the early universe.
%As the universe cools down, heavier dark fermion freezes out first and lighter $A_X$ later.

The relic density of the dark fermion is determined by the annihilation process 
$\psi\bar\psi \rightarrow A_X A_X$ at the time of thermal freezeout.
There is also another annihilation process $\psi\bar\psi \rightarrow A_X h_2$, 
but its contribution to the relic density is negligible.
The annihilation cross section $\times$ relative velocity is determined 
by the dark fermion mass $m_\psi$ and coupling $g_\psi \equiv g_X X_\psi$. 
Note that $g_\psi$ is independent of $g_X$ 
because the dark U(1) charge $X_\psi$ can be chosen as we want.
The relevant approximate formula for the process is given by \cite{Ma}
	\begin{eqnarray}
		\sigma(\psi\bar\psi \rightarrow A_X A_X) \times v_{rel} =
		 \frac{g_\psi^4}{16\pi m_\psi^2}.
	\end{eqnarray}
For $m_\psi = 100$ GeV and $g_\psi = 0.2$, it results in $\Omega_\psi h^2 = 0.12$.
Then, the dark fermion constitutes the dominant DM component with the correct DM relic density which we observe today.

Besides the DM fermion $\psi$, we consider a case that the dark gauge boson $A_X$ constitutes the subdominant DM component whose relic density is much smaller than the dominant one.
The annihilation process $A_X A_X \rightarrow h_2 h_2$ at the early universe determines the relic density of the dark gauge boson.
In this work, we will fix that $m_{h_2} = 0.8 \, m_{A_X}$ to guarantee no kinematic suppression for the process, and require that
the annihilation cross section $\times$ relative velocity is at least 10 times larger than the canonical value for the correct DM relic density in order to make the dark gauge boson $A_X$ be the subdominant DM.

%\sout{Since $A_X$ is stable but $h_2$ is unstable due to the unavoidable mixing with 
%the SM Higgs boson $h_1$, the chosen range of those light masses coincides with 
%the current observation of core-cusp anomaly in dwarf galaxies as discussed 
%in Refs.}\cite{Ma,Feng}.

With a light vector mediator $A_X$, the DM fermion self-interacting cross section 
would be enhanced at a small DM velocity by the Sommerfeld enhancement. 
For each combination of the dark fermion $\psi$ and the vector mediator $A_X$ mass, 
with the $g_\psi$ coupling value which is adjusted to give a correct relic density $\Omega_\psi h^2$, 
the Sommerfeld enhancement factor can be calculated.
It was shown that the DM self-interacting cross section could explain astrophysical observables at dwarf galaxy scale with the masses 
roughly in the following range \cite{Bringmann}:
\begin{equation} \label{eq:SIbound}
1\ \textrm{MeV} \lesssim m_{A_X} \lesssim 1\ \textrm{GeV}, \qquad  m_\psi \sim \mathcal{O}(100)\ \textrm{GeV} . 
\end{equation}
We study particle phenomenology and future discovery potential of the new physics model 
in this mass range of $m_{A_X}$ assuming $m_{A_X} \sim m_{h_2}$.  
Since $m_\psi$ and $g_\psi$ do not affect the particle phenomenology discussed in this work, 
one can easily choose $m_\psi$ and $g_\psi$ to satisfy the current relic density observation.

\subsection{Higgs decays}
\subsubsection{SM-like Higgs decays}
 
With the light new particles $(m_{h_2} \simeq m_{A_X} \ll m_{h_1}$), 
the SM-like Higgs $h_1$ has two additional decay modes: $h_1 \rightarrow h_2 h_2$ and $h_1 \rightarrow A_X A_X$ decays.
For $h_1 \rightarrow h_2 h_2$ decays, the corresponding decay rate is 
\begin{eqnarray}
	\Gamma(h_1 \rightarrow h_2 h_2)
%	= \frac{(2\,c_{122})^2}{32\pi m_{h_1}} \sqrt{1-\frac{4 m_{h_2}^2}{m_{h_1}^2}} \,
	\simeq \,\frac{1}{32 \pi m_{h_1}} (\lambda_{hs} v_h)^2.
\end{eqnarray}
For $h_1 \rightarrow A_X A_X$ decays, it is 
\begin{eqnarray}
	\Gamma(h_1 \rightarrow A_X A_X) 
%	&=& \frac{(2g_X m_{A_X} {\rm sin\theta_s})^2}{32\pi m_{h_1}} \sqrt{1-\frac{4 m_{A_X}^2}{m_{h_1}^2}}
%    \bigg(2+\frac{(m_{h_1}^2-2 m_{A_X}^2)^2}{4 m_A^4}\bigg) \nonumber\\
    \simeq   \frac{1}{32\pi m_{h_1}}  \bigg(g_X\, {\rm sin\theta_s}\,\frac{m_{h_1}^2}{m_{A_X}}\bigg)^2.
\end{eqnarray}
Because the scalar mixing angle $\theta_s$ in the limit of small mixing is given by
\begin{eqnarray}
	\theta_s \simeq \frac{\lambda_{hs} v_h v_s}{m_{h_1}^2},
\end{eqnarray}
we get following relations
\begin{eqnarray}
	\lambda_{hs} v_h \simeq \theta_s \frac{m_{h_1}^2}{v_s} \simeq {\rm sin}\theta_s\, g_X \frac{m_{h_1}^2}{m_{A_X}}.
\end{eqnarray}
Then, the two decay rates are almost the same to each other,
\begin{eqnarray}
	\Gamma(h_1 \rightarrow h_2 h_2) \simeq \Gamma(h_1 \rightarrow A_X A_X).
\end{eqnarray}
The decay rate for the exotic Higgs decay channels can be written in terms of the new physics parameters as
\begin{eqnarray}
	\Gamma(h_1 \rightarrow {\rm exo}) \equiv \Gamma(h_1 \rightarrow h_2 h_2) + \Gamma(h_1 \rightarrow A_X A_X) 
	\simeq \frac{1}{16\pi m_{h_1}} \bigg(g_X\, {\rm sin\theta_s}\,\frac{m_{h_1}^2}{m_{A_X}} \bigg)^2.
\end{eqnarray}

The decay products of the exotic Higgs decays would appear invisible to the detector. Direct searches for invisible Higgs decays 
have been carried out with the ATLAS and the CMS detectors at the Large Hadron Collider (LHC).
The recent results of the upper limit on the branching ratio of the invisible Higgs decays correspond to
about 0.1 at 95$\%$ confidence level\cite{ATLAS}. Here, we will demand that the branching ratio of the exotic Higgs decay, 
Br$(h_1 \rightarrow {\rm exo})$ is less than 0.1 for searching allowed parameter spaces of the two-component DM model.

%\vspace{1.0cm}
\subsubsection{Singlet-like Higgs decays}

The decay pattern of the singlet-like Higgs $h_2$ is exactly the same as the SM-like Higgs case of the same mass. 
But the decay rates are suppressed by a overall factor of $\rm sin^2\theta_s$, compared to the SM-like Higgs case.
For a light $h_2$ of order 10 MeV in mass, its dominant decay channel is to an electron-positron pair and the corresponding
decay rate is 
\begin{eqnarray}
	\Gamma(h_2 \rightarrow e^+e^-) =
   \frac{m_{h_2}}{8\pi}\bigg(\frac{m_e}{v_h}\,{\rm sin\theta_s}\bigg)^2\bigg(1-\frac{4 m_e^2}{m_{h_2}^2}\bigg)^{3/2}.
\end{eqnarray}

It would give a rather large lifetime for a light singlet-like Higgs if $\rm sin\theta_s$ is very small. 
For instance, when $m_{h_2} = 8$ MeV and $\rm sin\theta_s = 2 \times 10^{-5}$,
$h_2$ lifetime is about 1.2 seconds. Such a large lifetime might spoil the successful Big Bang Nucleosynthesis (BBN).
For the allowed model parameter space, we will require that $h_2$ lifetime should be less than 1 second.

\subsection{Heavy meson decays}

We may use rare $B$ decays to search for new physics signals or constrain the new physics model parameters.
We focus on $B \rightarrow A_X A_X$ decays and $B \rightarrow K A_X A_X$ decays, which would appear
as $B \rightarrow {\rm invisible}$ and $B \rightarrow K + {\rm invisible}$ to detector, respectively.
The effective Lagrangian relevant for the rare $B$ decays\cite{Batell} is
\begin{eqnarray}
	\mathcal{L} = \frac{3\sqrt{2}G_F m_t^2 V_{tq}^* V_{tb}}{16\pi^2} 
	\frac{m_b}{v_h}\, h\, \bar{q}_L b_R + \textrm{h.c.} ,
\end{eqnarray}
where $h= h_1\,{\rm cos}\theta_s - h_2\,{\rm sin}\theta_s$, $G_F$ is the Fermi constant, and $V_{tq}$ is a relevant CKM matrix element for the process. 

\subsubsection{$B^0 \rightarrow A_X A_X$ decays and $B^0 \rightarrow {\rm invisible}$}

The matrix element squared for $B^0$ meson decays, $B^0(p) \rightarrow A_X(k_1) A_X(k_2)$, is
\begin{eqnarray}
 \sum_{\textrm{spin}} |\mathcal{M}|^2 = \frac{|c_0|^2}{4} {\rm sin^2 2\theta_s}\, g_X^2 m_{A_X}^2 \,f_B^2
	\bigg(\frac{m_B^2}{m_b+m_q}\bigg)^2 
	\bigg(\frac{1}{m_B^2-m_{h_1}^2} - \frac{1}{m_B^2-m_{h_2}^2} \bigg)^2 
	\bigg(2 + \frac{(k_1\cdot k_2)^2}{m_{A_X}^4}\bigg),
\end{eqnarray}	
where the Wilson coefficient $c_0\equiv \frac{3\sqrt{2}G_F m_t^2 V_{tq}^* V_{tb}}{16\pi^2} \frac{m_b}{v_h}$, 
and the $B$ meson decay constant $f_B$ is defined as
%\begin{eqnarray}
%	<0|\bar{q}\gamma_\mu\gamma_5 b|B> = i f_B p_\mu \,\,\,\, {\rm or} \,\,\,\, 
%	(m_b+m_q) <0|\bar{q}i\gamma_5 b|B> = m_B^2 f_B.
%\end{eqnarray}	
\begin{equation}
	(m_b+m_q) <0|\bar{q}i\gamma_5 b|B> = m_B^2 f_B.
\end{equation}
For a small $m_{A_X} \simeq m_{h_2} \simeq 10$ MeV, ${\rm sin\theta_s = 0.0002}$, $g_X = 0.005$, and $f_B=190$MeV,
the branching ratio of the decays $B^0 \rightarrow A_X A_X$ is 
\begin{eqnarray}
	 {\rm Br}(B^0 \to A_X A_X)  \simeq 2.8 \times 10^{-11} 
	\bigg(\frac{{\rm sin}\,2\theta_s}{0.0004}\bigg)^2
	\bigg(\frac{g_X}{0.005}\bigg)^2 \bigg(\frac{0.01\, {\rm GeV}}{m_{A_X}}\bigg)^2.
\end{eqnarray}

At present, BaBar collaboration has established the most stringent upper limit  of $2.4 \times 10^{-5}$ at the 90$\%$ confidence level for the branching ratio of $B^0 \rightarrow {\rm invisible}$\cite{Babar}.
We require ${\rm Br}(B^0\rightarrow A_X A_X)$ should be smaller than $2.4 \times 10^{-5}$ for constraining
the new physics model parameters.

\subsubsection{$B \rightarrow K A_X A_X$, $K h_2$ decays and $B \rightarrow K + {\rm invisible}$}

For the three-body $B$ decays, $B(p_B) \rightarrow K(p_K) A_X(k_1) A_X(k_2)$, with $q\equiv p_B-p_K$,
the matrix element is given by
\begin{eqnarray}
	i\mathcal{M} = ic_0 <K|\bar{s}\frac{1}{2}(1+\gamma_5)b |B> {\rm cos\theta_s}\,{\rm sin\theta_s} \,
	\bigg(\frac{i}{q^2-m_1^2}-\frac{i}{q^2-m_2^2} \bigg) \,2g_X^2 v_s\, \epsilon_\mu^*(k_1) \epsilon^{\mu*}(k_2),
\end{eqnarray}
where the hadronic matrix element is related to a hadronic form factor as follows,
\begin{eqnarray}
	<K|\bar{s}\frac{1}{2}(1+\gamma_5)b|B> = 
	\frac{1}{2}\frac{q^\mu}{m_b}<K|\bar{s}\gamma_\mu b|B> =
%	\frac{1}{2}\frac{q^\mu}{m_b} <K(p_K)|\bar{s}\gamma_\mu b|B(p_B)> = 
	\frac{1}{2 m_b} (m_B^2 - m_K^2) f_0^K(q^2). 
\end{eqnarray} 

Then, the decay rate for the process is obtained as
\begin{eqnarray}	
	\Gamma =\frac{1}{512\pi^3 m_B^3}  \int_{4 m_{A_X}^2}^{(m_B-m_K)^2} dq^2 \sum_{\textrm{spin}} |\mathcal{M}|^2 
	\sqrt{q^2 - 4m_{A_X}^2} \sqrt{\frac{(m_B^2 - q^2 - m_K^2)^2}{q^2} - 4 m_K^2},	
\end{eqnarray}
where the matrix element squared is 
\begin{eqnarray}
	\sum_{\textrm{spin}} |\mathcal{M}|^2 = \frac{|c_0|^2}{4} {\rm sin^2 2\theta_s} \, g_X^2 m_{A_X}^2 
	\frac{(m_B^2-m_K^2)^2}{m_b^2} (f_0^K(q^2))^2 
	\bigg(\frac{1}{q^2-m_1^2}-\frac{1}{q^2-m_2^2} \bigg)^2 
	\bigg(2 + \frac{(k_1\cdot k_2)^2}{m_A^4}\bigg),  	
\end{eqnarray}
with $k_1 \cdot k_2 = \frac{1}{2} (q^2-2m_{A_X}^2)$. 

For numerical calculation, we adopt the form factor $f_0^K(q^2)$ as follows\cite{Ball}: 
%(Table 3, Eq.(32) in hep-ph/0406232)
\begin{eqnarray}
	f_0^K(q^2) = \frac{0.33}{1-q^2/37.46 \,({\rm GeV}^2)}.
\end{eqnarray}
For a small $m_{A_X} = 10$ MeV, $m_{h_2} = 0.8\, m_{A_X}$, ${\rm sin\theta_s = 0.0002}$, $g_X = 0.005$,
the branching ratio of the decay $B \rightarrow K A_X A_X$ is 
\begin{eqnarray}
	{\rm Br}(B^- \rightarrow K^- A_X A_X) \simeq 3.3 \times 10^{-10}
	\bigg(\frac{{\rm sin}\,2\theta_s}{0.0004}\bigg)^2
\bigg(\frac{g_X}{0.005}\bigg)^2 \bigg(\frac{0.01\, {\rm GeV}}{m_{A_X}}\bigg)^2.
\end{eqnarray}
For $m_{A_X} = 1$ GeV, $m_{h_2} = 0.8\, m_{A_X}$, and with the same values for ${\rm sin\theta_s = 0.0002}$, $g_X = 0.005$,
the branching ratio is ${\rm Br}(B \rightarrow K A_X A_X) \simeq 1.8\times 10^{-14}$.

The branching ratio of $B^+ \rightarrow K^+ \nu\bar\nu$ is predicted to be $(4.6\pm 0.5)\times 10^{-6}$
in the SM.
The search for the decays $B^+ \rightarrow K^+ \nu\bar\nu$ was performed at the Belle II experiment. 
An upper limit on the branching ratio Br($B^+ \to K^+\nu\bar{\nu}$)
of $4.1 \times 10^{-5}$ is set at the 90$\%$ confidence level\cite{BelleII}. 
We require that Br($B^+ \rightarrow K^+ + {\rm invisible}$) should be less than $4.1\times 10^{-5}$.

If $h_2$ lifetime is large, it would be invisible to detector.
In that case, the process $B \rightarrow K h_2$ also contributes to $B \rightarrow K + {\rm invisible}$ decays.
For $B(p_B) \rightarrow K(p_K) h_2(q)$ decays, the corresponding matrix element is
\begin{eqnarray}
	i\mathcal{M} = ic_0 <K|\bar{s}\frac{1}{2}(1+\gamma_5)b|B> {\rm sin\theta_s} = 
	ic_0\,\frac{1}{2m_b} (m_B^2-m_K^2) f_0^K (q^2) \,{\rm sin\theta_s},
\end{eqnarray}
and the decay rate is
\begin{eqnarray}
	\Gamma = \frac{|\bf{q}|}{8\pi m_B^2} |\mathcal{M}|^2,
\end{eqnarray}
where $\bf{q}$ is the momentum of the final state particles in the $B$ meson rest frame, 
\begin{eqnarray}
	|{\bf{q}}| = |{\bf{p}}_K| = \frac{\sqrt{(m_B^2 - (m_K+m_{h_2})^2)(m_B^2-(m_K-m_{h_2})^2)}}{2 m_B} .
\end{eqnarray}
For $m_{h_2}$ = 10 MeV, and $\rm sin\theta_s$ = 0.0002, the branching ratio is about
\begin{eqnarray}
	{\rm Br}(B^- \rightarrow K^- h_2) \sim 1.7 \times 10^{-8}\, \bigg(\frac{\rm sin\theta_s}{0.0002}\bigg)^2.
\end{eqnarray}
We will consider $B \rightarrow K h_2$ decays as $B \rightarrow K + {\rm invisible}$ events 
if $m_{h_2} < 2 m_\mu$. 

\subsection{Allowed parameter space and future prospects}

We adopt free model parameters as follows: the singlet-like scalar mass $m_{h_2}$, 
the scalar mixing angle ${\rm sin}\theta_s$, the dark gauge boson mass $m_{A_X}$, 
and the dark gauge coupling $g_X$, in addition to the dark fermion mass $m_\psi$ and 
the dark U(1) charge $X_\psi$ of the dark fermion.
Here, the dark fermion mass $m_\psi$ and charge $X_\psi$ can be freely adjusted to achieve 
a right amount of the dark matter relic density as the dominant component of the dark matter.
We fix the dark gauge boson mass as $m_{A_X} = m_{h_2}/0.8$ to have a sufficient annihilation cross section
for $A_X A_X \rightarrow h_2 h_2$, in order to have a small enough relic density as the subdominant component
of the dark matter.
Then, only 3 free parameters remain: $m_{h_2}$, ${\rm sin\theta_s}$, and $g_X$.

Now we investigate the allowed model parameter space which satisfy various experimental and phenomenological constraints, 
and study the possibility for finding new physics signals from future experiments.
We show the experimental and phenomenological constraints and 
the sensitivity of SHiP\cite{SHiP}, DUNE\cite{DUNE}, and ILC\cite{ILC} experiments 
on the parameter plane $(m_{h_2}, {\rm sin}\theta)$ for the given $g_X$ values.

%\begin{figure}[!hbt]
%	%	\centering
%	\includegraphics[height=6cm]{invhiggs1.png} \,
%	\includegraphics[height=6cm]{invhiggs2.png} \\
%	\includegraphics[height=6cm]{invhiggs3.png} \,
%	\includegraphics[height=6cm]{invhiggs4.png}
%	\caption{
%	  Model parameter space of ($m_{h_2}, {\rm sin\theta_s}$) allowed by the current experimental constraints
%	  and phenomenological bounds is shown for given $g_X$ values. The four figures correspond to
%	  four different $g_X$ values, i.e., $g_X = 0.005, 0.01, 0.05$, and 0.1, respectively.
%	  Also are shown the sensitivity regions of the future experiments
%	  such as the SHiP, DUNE and ILC experiments.
%	}
%\end{figure}
%%%%%%%%%%%%%%%%%%%%%%%%%%%%%%%%%%%%%%%%%%%%%%%%%%
\begin{figure}[!hbt]
	\centering%
	\subfigure[$g_X = 0.005$ ]{\label{invhiggs1} %
		\includegraphics[width=7.4cm]{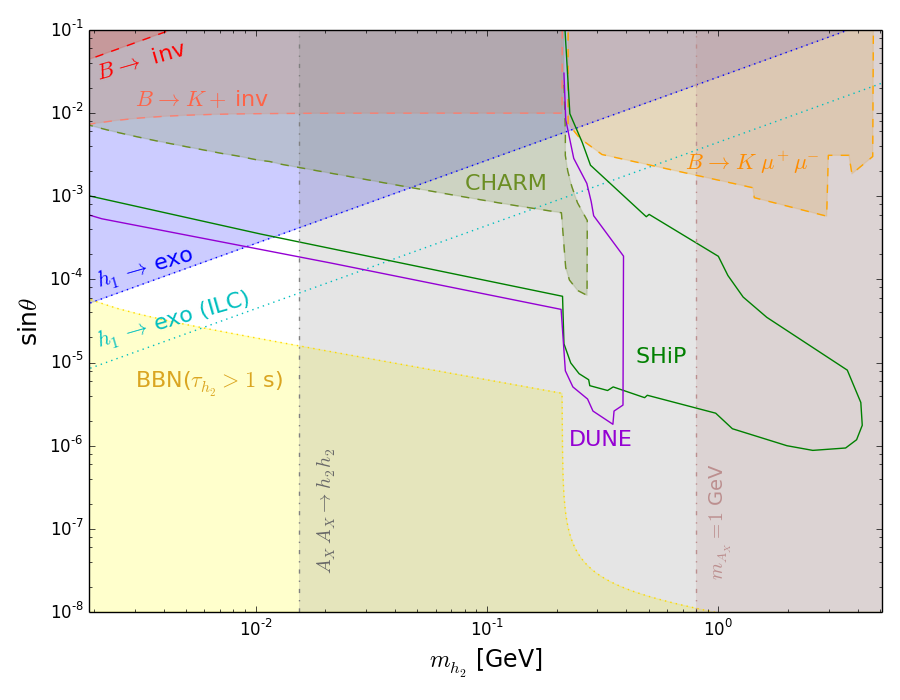}} \
	\subfigure[$g_X = 0.01$ ]{\label{invhiggs2} %
		\includegraphics[width=7.4cm]{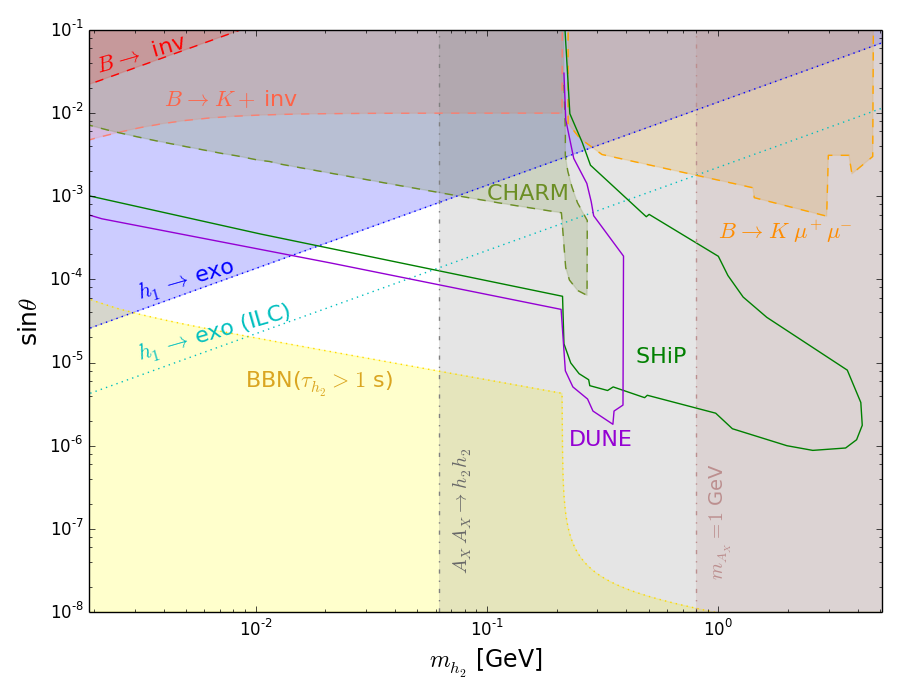}} \\   
	\subfigure[$g_X = 0.05$ ]{\label{invhiggs3} %
		\includegraphics[width=7.4cm]{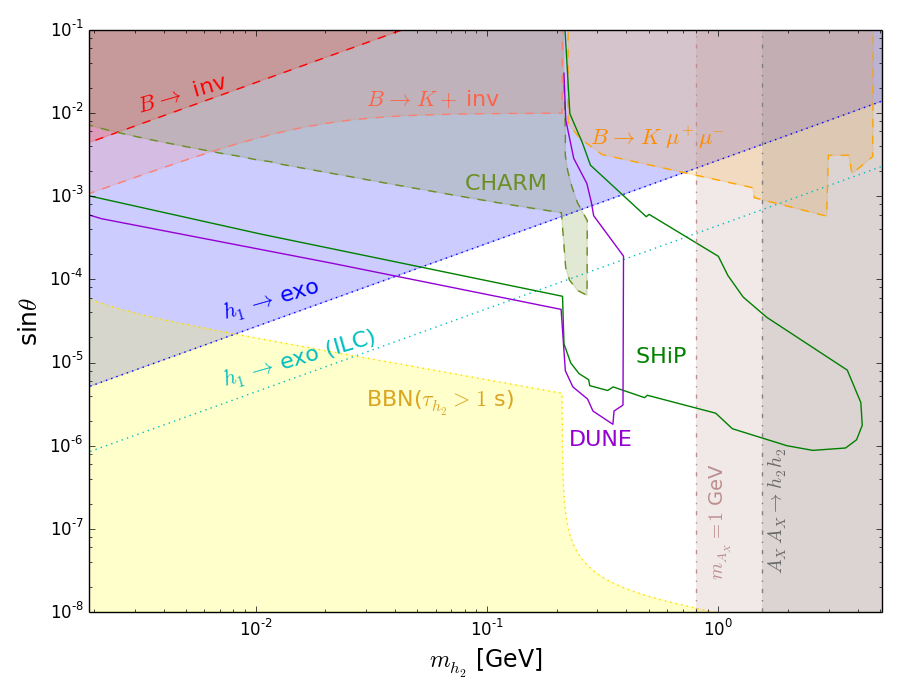}} \
	\subfigure[$g_X = 0.1$ ]{\label{invhiggs4} %
		\includegraphics[width=7.4cm]{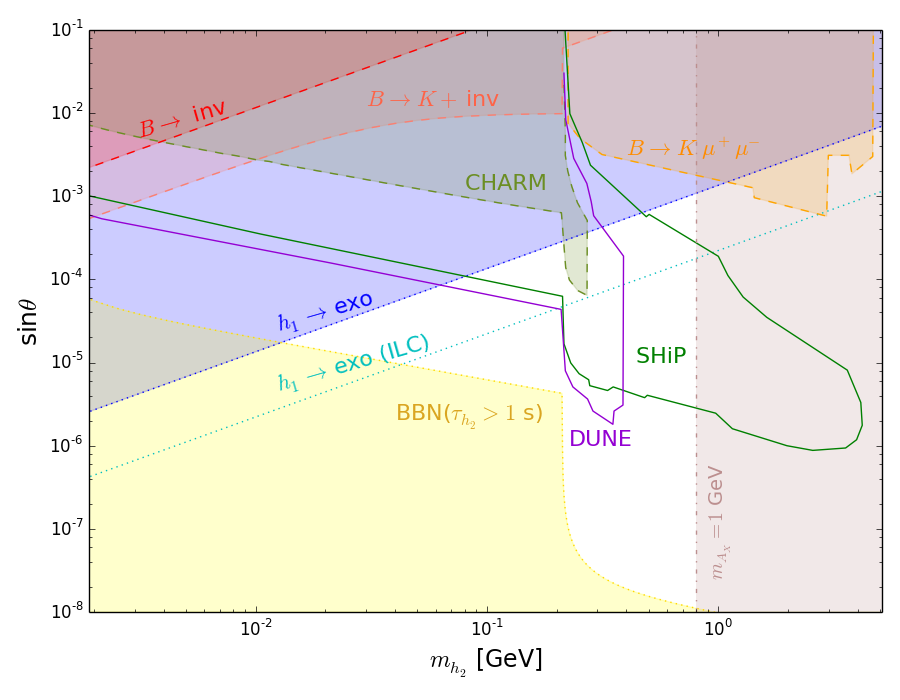}}     
	\caption{Model parameter space of ($m_{h_2}, {\rm sin\theta_s}$) allowed by the current experimental constraints
		and phenomenological bounds is shown for given $g_X$ values. The four figures correspond to
		four different $g_X$ values, i.e., $g_X = 0.005, 0.01, 0.05$, and 0.1, respectively.
		Also are shown the sensitivity regions of the future experiments
		such as the SHiP, DUNE and ILC experiments.
	   The vertical lines indicate the upper limits on $m_{h_2}$ obtained from Eq.~(\ref{eq:sigma_Ax})
	  	and from Eq.~(\ref{eq:SIbound}) for $m_{h_2} = 0.8 m_{A_X}$, respectively.  } 
	\label{fig:invhiggs}
\end{figure}
%%%%%%%%%%%%%%%%%%%%%%%%%%%%%%%%%%%%%%%%%%%%%%%%%%

Fig.~\ref{invhiggs1} shows ($m_{h_2}, \rm sin\theta$) parameter space for $g_X = 0.005$.
On the figure, $h_1 \rightarrow \rm exo$ line corresponds to the contour line 
for the branching ratio of the exotic Higgs decays, Br$(h_1 \rightarrow {\rm exo}) = 10\%$.
The region above the line is excluded by the current upper limit on Br$(h_1 \rightarrow {\rm exo})$ at the LHC.
The (yellow-colored) BBN($\tau_{h_2} > 1$s) region on the figure corresponds to the parameter space where the lifetime of $h_2$ exceeding 1 second is predicted. 
The parameter region of BBN($\tau_{h_2} > 1$s) should be avoided for the successful BBN phenomenology.
The lines of $B \rightarrow {\rm inv}$ and $B \rightarrow K + {\rm inv}$ correspond to
the contour lines for Br($B \rightarrow {\rm invisible}) = 2.4 \times 10^{-5}$ and 
Br($B \rightarrow K + {\rm invisible}) = 4.1 \times 10^{-5}$, respectively.
The (shaded) regions above the lines are excluded by the current upper limits on the branching ratios at the B-factories.
 
For annihilation process $A_X A_X \rightarrow h_2 h_2$, the resulting cross section $\times$ relative velocity is
proportional to $g_X^4 / m_{A_X}^2$ for a fixed mass ratio $r = m_{h_2}^2 / m_{A_X}^2$. 
With $r=0.64$, we have
\begin{eqnarray} \label{eq:sigma_Ax}
	\sigma (A_X A_X \rightarrow h_2 h_2) \, v_{rel} = 1.1 \times 10^{-24}  \, 
	\bigg(\frac{g_X}{0.005}\bigg)^4 \bigg(\frac{0.01\,{\rm GeV}}{m_{A_X}}\bigg)^2 {\rm cm^3 \,s^{-1}},
\end{eqnarray}
which is 38 times the canonical value of $3 \times 10^{-26}\, {\rm cm^3\,s^{-1}}$ for having 
the right amount of DM relic density of the Universe. Here, we require that the cross section $\times$ relative velocity
should be at least 10 times larger than the canonical value in order to get a much smaller relic density for
the subdominant DM $A_X$ than the measured DM density of the Universe, 
 in such a way that the relic density and the DM self-interaction cross section of $\psi$ are exclusively
	determined by $m_\psi$ and $g_\psi$.   
For $g_X = 0.005$, it means that $m_{A_X}$ should be smaller than about 0.2 GeV. 
In turn, it implies that $m_{h_2}$ should be smaller than about $0.16$ GeV, for the fixed $r=0.64$.
In Fig.~\ref{fig:invhiggs}, 
the vertical lines of $A_X A_X \rightarrow h_2 h_2$ and $m_{A_X} = 1$ GeV
indicate the upper limits on the dark scalar boson mass obtained from Eq.~(\ref{eq:sigma_Ax})
and from Eq.~(\ref{eq:SIbound}) for $m_{h_2} = 0.8 m_{A_X}$, respectively. 

After imposing current experimental bounds and phenomenological constraints, 
only small triangular shaped region remains allowed, 
where 2 MeV $\lesssim m_{h_2} \lesssim$ 16 MeV and $2 \times 10^{-5} \lesssim {\rm sin\theta_s} \lesssim 3 \times 10^{-4}$. 
On that figure, we also show the projected sensitivity of the SHiP and DUNE experiments. 
It indicates that some small part of the allowed parameter region
with $m_{h_2} \sim 10$ MeV and $\rm sin\theta_s \sim 3 \times 10^{-4}$ would be probed 
by the future experiments. We also notice that, in the allowed parameter space, 
the branching ratio of the exotic Higgs decay is about 0.025 $\% \lesssim$ Br($h_1 \rightarrow {\rm exo}) \lesssim 10 \%$,
which might be probed by the future leptonic collider such as ILC whose sensitivity on the 
branching ratio of the invisible Higgs decay is expected to reach $0.3 \%$ level\cite{ILC}.
We also denote the corresponding contour lines for the ILC reach on the figures.  

Figs.~\ref{invhiggs1}, \ref{invhiggs2}, and \ref{invhiggs3} show the same parameter spaces ($m_{h_2}, {\rm sin\theta_s}$) 
for $g_X = 0.01, 0.05, 0.1$, respectively.
By increasing the $g_X$ values, the $h_1 \rightarrow {\rm exo}$ contour lines get lowered and therefore 
more parameter spaces are ruled out
because the branching ratio of the exotic Higgs decays is proportional to $(g_X\, {\rm sin\theta_s}/m_{A_X})^2$.
The contour lines for the $B \rightarrow \rm{inv}$ and $B \rightarrow K + {\rm inv}$ decays get also lowered
similarly to the $h_1 \rightarrow {\rm exo}$ case.
%because the branching ratios of $B \rightarrow A_X A_X$ and $B \rightarrow K A_X A_X$ decays
%are all proportional to $(g_X\, {\rm sin\theta_s}/m_{A_X})^2$}}. 
However, the constraints from the exotic Higgs decays are always more stringent than the one from the $B$ decays. 
For the $\sigma(A_X A_X \rightarrow h_2 h_2) \, v_{rel}$, it is proportional to $g_X^4/m_{A_X}^2$.
Therefore, the corresponding contour lines of the lower limits on $\sigma(A_X A_X \rightarrow h_2 h_2) \,v_{rel}$ 
are shifted to larger $m_{A_X}$ values when $g_X$ increases. 
Hence, the allowed parameter spaces get broader by increasing $g_X$ value. 
On the other hand, the lifetime of the singlet-like Higgs $h_2$ does not depend on $g_X$.
Therefore, the excluded regions by the upper limit of $h_2$ lifetime are all the same for different $g_X$.  

For $g_X=0.005$, the SHiP and DUNE sensitivity regions barely touch the allowed parameter space of the model.
On the other hand, for larger $g_X$, rather large portion of the allowed parameter space 
would be explored by the SHiP and DUNE experiments.
Also the future ILC experiments would explore a fair amount of the allowed parameter space
through the exotic Higgs decays for all $g_X$ values.

\section{Conclusions}

We studied phenomenology of a two-component DM model, which would provide a solution for the small scale problems
while avoiding the strong constraints from the direct DM detection experiments and the CMB observations.
A new Dirac fermion introduced as a main component of DM in the Universe, 
which explains the observed relic density, does not directly interact with
the SM particles, thus avoiding strong constraints from the direct DM detection experiments.
On the other hand, the main DM component is self-interacting through a light dark gauge boson 
which plays a subdominant DM and connects with the SM sector via Higgs portal interaction. 
The self-interacting DM would solve the small scale problem such as the core-cusp problem.

We investigated the model parameter space allowed by the current experimental constraints
from the Higgs and $B$ decays and the phenomenological bounds from the successful BBN and 
from the requirement of a large enough annihilation cross section of the dark gauge boson to make it a subdominant DM. 
We showed the allowed parameter space on the ($m_{h_2}, \,\rm {sin\theta_s}$) plane for various $g_X$ values.
For $g_X=0.005$, the region of $2\,{\rm MeV} \lesssim m_{h_2} \lesssim 16\, {\rm MeV}$ and 
$2\times 10^{-5} \lesssim {\rm sin\theta_s} \lesssim 3 \times 10^{-4}$ are allowed by the current experimental constraints
and the phenomenological requirements. By increasing $g_X$, both the minimum and the maximum 
of the allowed $m_{h_2}$ values are shifted to larger values and the allowed region of $\rm \sin\theta_s$ becomes broader.

We also discussed the sensitivity of the future experiments such as the SHiP, DUNE, and ILC 
for the obtained allowed parameter space.
It turns out that large portion of the allowed parameter space in this model could be explored by the future experiments.

\section*{Acknowledgments}
This work was supported by the Basic Science Research Program through the National Research Foundation of Korea (NRF)
funded by the Ministry of Education under the Grant No. NRF-2020R1I1A1A01072816 (ShN),
and also funded by the Ministry of Science and ICT under the Grant Nos. 
NRF-2021R1F1A1061717(YGK), NRF-2021R1A2C2011003 (KYL), and NRF-2020R1A2C3009918 (ShN).


\begin{thebibliography}{99}

%\bibliographystyle{rnsrt}
%\bibliography{rnsrt}

\bibitem{Bertone}
  G.~Bertone, D.~Hooper and J.~Silk,
  ``Particle Dark Matter: Evidence, Candidates and Constraints'',
  Phys.\ Rep.\ {\bf 405}, 279 (2005)
  [arXiv:hep-ph/0404175].

\bibitem{Bauer}
M.~Bauer and T.~Plehn,
``Yet Another Introduction to Dark Matter: The Particle Physics Approach,''
Lect. Notes Phys. \textbf{959}, pp. (2019)
%doi:10.1007/978-3-030-16234-4
[arXiv:1705.01987 [hep-ph]].

\bibitem{Profumo}
S.~Profumo, L.~Giani and O.~F.~Piattella,
``An Introduction to Particle Dark Matter,''
Universe \textbf{5}, no.10, 213 (2019)
%doi:10.3390/universe5100213
[arXiv:1910.05610 [hep-ph]].

\bibitem{LUX} 
D.~S.~Akerib {\it et al.} [LUX Collaboration],
``Results from a Search for Dark Matter in the Complete LUX Exposure'',
Phys.\ Rev.\ Lett. {\bf 118}, 021303 (2017)
[arXiv:1608.07648 [astro-ph.CO]]. 

\bibitem{PandaX}
X.~Cui {\it et al.} [PandaX-II Collaboration],
``Dark Matter Results from 54-Ton-Day Exposure of PandaX-II Experiment'',
Phys.\ Rev.\ Lett. {\bf 119}, 181302 (2017)
[arXiv:1708.06917 [astro-ph.CO]].

\bibitem{XENON1T}
  E.~Aprile {\it et al.} [XENON Collaboration],
  ``Dark Matter Search Results from a Ton-Year Exposure of XENON1T'',
  Phys.\ Rev.\ Lett. \ {\bf 121}, 111302 (2018)
  [arXiv:1805.12562 [astro-ph.CO]].
  
\bibitem{Monroe}
  J.~Monroe and P.~Fisher, 
  ``Neutrino Backgrounds to Dark Matter Searches'',
  Phys.\ Rev.\ D {\bf 76}, 033007 (2007)
  [arXiv:0706.3019 [astro-ph]].
  
\bibitem{Vergados}  
J.~D.~Vergados and H.~Ejiri,
``Can Solar Neutrinos be a Serious Background in Direct Dark Matter Searches?'',
Nucl.\ Phys.\ B {\bf 804}, 144 (2008)
[arXiv:0805.2583 [hep-ph]].

\bibitem{Gutlein}
A.~Gutlein {\it et al.},
 ``Solar and atmospheric neutrinos: Background sources for the direct dark matter searches'',
Astropart.\ Phys. {\bf 34}, 90 (2010)
[arXiv:1003.5530 [hep-ph]].

\bibitem{Billard}
J.~Billard, L.~Strigari and E.~Figueroa-Feliciano,
``Implication of neutrino backgrounds on the reach of next generation dark matter direct detection experiments'',
Phys.\ Rev.\ D {\bf 89}, 023524 (2014)
[arXiv:1307.5458 [hep-ph]].

\bibitem{OHare}
C.~A.~J.~O'Hare,
``Can we overcome the neutrino floor at high masses?'',
Phys.\ Rev.\ D {\bf 102}, 063024 (2020)
[arXiv:2002.07499 [astro-ph.CO]].

\bibitem{Tulin}
  S.~Tulin and H.-B.~Yu,
  ``Dark Matter Self-interactions and Small Scale Structure'',
  Phys.\ Rept.\ {\bf 730}, 1 (2018)
  [arXiv:1705.02358 [hep-ph]].

\bibitem{Dubinski}
J.~Dubinski and R.~G.~Carlberg,
``The Structure of Cold Dark Matter Halos'',
Astrophys.\ J.\ {\bf 378}, 496 (1991).
  
\bibitem{NFW1}
J.~F.~Navarro, C.~S.~Frenk and S.~D.M.~White,
``The Structure of Cold Dark Matter Halos'',
Astrophys.\ J.\ {\bf 462}, 563 (1996)
[arXiv:astro-ph/9508025].

\bibitem{NFW2}
J.~F.~Navarro, C.~S.~Frenk and S.~D.M.~White,
``A Universal Density Profile from Hierarchical Clustering'',
Astrophys.\ J.\ {\bf 490}, 493 (1997)
[arXiv:astro-ph/9611107].
  
\bibitem{Flores}  
R.~A.~Flores and J.~R.~Primack,
``Observational and Theoretical Constraints on Singular Dark Matter Halos'', Astrophys.\ J.\ {\bf 427}, L1 (1994)
[arXiv:astro-ph/9402004]. 
  
\bibitem{Moore}  
B.~Moore,
``Evidence against dissipationless dark matter from observations of galaxy halos'',
Nature {\bf 370}, 629 (1994).

\bibitem{Burkert}
A.~Burkert,
``The Structure of Dark Matter Haloes in Dwarf Galaxies'',
Astrophys.\ J.\ Lett. {\bf 447}, L25 (1995)
[arXiv:astro-ph/9504041].

\bibitem{McGaugh}
S.~S.~McGaugh, W.~J.~G.~de Blok, 
``Testing the Dark Matter Hypothesis with Low Surface Brightness Galaxies and Other Evidence'',
Astrophys.\ J.\ {\bf 499}, 41 (1998)
[arXiv:astro-ph/9801123].

\bibitem{Spergel}
D.~N.~Spergel and P.~J.~Steinhardt,
``Observational evidence for self-interacting cold dark matter'',
Phys.\ Rev.\ Lett. {\bf 84}, 3760 (2000)
[arXiv:astro-ph/9909386].

\bibitem{Galli}
S.~Galli, F.~Iocco, G.~Bertone and A.~Melchiorri,
``CMB constraints on dark matter models with large annihilation cross section'',
Phys.\ Rev.\ D {\bf 80}, 023050 (2009)
[arXiv:0905.0003 [astro-ph.CO]].

\bibitem{Bringmann}
T.~Bringmann, F.~Kahlhoefer and K.~Schmidt-Hoberg,
``Strong Constraints on Self-Interacting Dark Matter with Light Mediators'',
Phys.\ Rev.\ Lett. {\bf 118}, 141802 (2017)
[arXiv:1612.00845 [hep-ph]].
  
\bibitem{Ma}
  E.~Ma,
  ``Inception of self-interacting dark matter with dark charge conjugation symmetry'',
 Phys.\ Lett.\ B {\bf 772}, 442-445 (2017)
 [arXiv:1704.04666 [hep-ph]].
 
\bibitem{ATLAS}
The ATLAS collaboration,
``Combination of searches for invisible Higgs boson decays with the ATLAS experiment'', ATLAS-CONF-2020-052.

\bibitem{Batell}
B.~Batell, M.~Pospelov and A.~Ritz,
``Multi-lepton Signatures of a Hidden Sector in Rare B Decays'',
Phys.\ Rev.\ D {\bf 83}, 054005 (2011)
[arXiv:0911.4938 [hep-ph]].

\bibitem{Babar}
J.~P.~Lees {\it et al.} [BABAR Collaboration],
``Improved limits on $B^0$ decays to invisible (+$\gamma$) final states'',
Phys.\ Rev.\ D {\bf 86}, 051105(R) (2012)
[arXiv:1206.2543 [hep-ex]].
 
\bibitem{Ball}
P.~Ball and R.~Zwicky,
``New Results on $B \rightarrow \pi, K, \eta$ Decay Formfactors from Light-Cone Sum Rules'',
Phys.\ Rev.\ D {\bf 71}, 014015 (2005)
[arXiv:hep-ph/0406232].

\bibitem{BelleII}
F.~Abudinen {\it et al.} [Belle II Collaboration], 
``Search for $B^+ \rightarrow K^+\nu\bar{\nu}$ decays using an inclusive tagging method at Belle II'',
Phys.\ Rev.\ Lett. {\bf 127}, 181802 (2021)
[arXiv:2104.12624 [hep-ex]].

\bibitem{SHiP}
S.~Alekhin {\it et al.},
``A facility to Search for Hidden Particles at the CERN SPS: the SHiP physics case'',
Rep.\ Prog.\ Phys. {\bf 79}, 124201 (2016)
[arXiv:1504.04855 [hep-ph]].

\bibitem{DUNE}
J.~M.~Berryman {\it et al.}, 
``Searches for decays of new particles in the DUNE Multi-Purpose near Detector'',
JHEP {\bf 02}, 174 (2020)
[arXiv:1912.07622 [hep-ph]].

\bibitem{ILC}
T.~Barklow {\it et al.},
``Improved Formalism for Precision Higgs Coupling Fits'',
Phys.\ Rev.\ D {\bf 97}, 053003 (2018)
[arXiv:1708.08912 [hep-ph]].

\end{thebibliography}
\end{document}